# Molten pool characteristics of a nickel-titanium shape memory alloy for directed energy deposition

Shiming Gao[1], Yuncong Feng[1], Jianjian Wang[2], Mian Qin[1], O.P Bodunde[1], Wei-Hsin Liao[1,*], Ping Guo[2,*]

[1]Department of Mechanical and Automation Engineering, The Chinese University of Hong Kong, Shatin, Hong Kong, China

[2]Department of Mechanical Engineering, Northwestern University, Evanston, IL, USA

*Corresponding authors: Wei-Hsin Liao: whliao@cuhk.edu.hk

Ping Guo: ping.guo@northwestern.edu

**Abstract**

Fabrication of nickel-titanium shape memory alloy through additive manufacturing has attracted increasing interest due to its advantages of flexible manufacturing capability, low-cost customization, and minimal defects. The process parameters in directed energy deposition (DED) have a crucial impact on its molten pool characteristics (geometry, microstructure, etc.), thus influencing the final properties of shape memory effect and pseudoelasticity. In this paper, a three-dimensional numerical model considering heat transfer, phase change, and fluid flow has been developed to simulate the cladding geometry, melt pool depth, and deposition rate. The experimental and simulated results indicated that laser power plays a critical role in determining the melt pool width and deposition rate while scan speed and powder feed rate have less effect on cladding geometry and deposition rate. The fluid velocity has a huge influence on the distribution of elements in the molten pool. The temperature gradient $G$, solidification rate $R$, as well as shape factor $G/R$ were calculated to illustrate the underlying mechanisms of grain structure evolution. The grain morphology distribution of cross-section from the experimental samples agreed well with the simulation results. The model reported in this paper is expected to shed light on the optimization of the deposition process and grain structure prediction.

## 1. Introduction

Technology is moving towards 'smart' systems with intelligent and adaptive features and functions. These distinct properties boost the development of various kinds of 'smart' materials (e.g. shape memory polymer, shape memory alloy, etc.). Near equiatomic nickel-titanium shape memory alloys (SMAs), as a typical representative, appear in a B2 structured austenite phase, a B19' structured martensite phase or a rhombohedral R-phase, depending on the thermal or mechanical conditions [1]. The martensitic and reverse-martensitic transformation between the abovementioned phases enables SMAs to regain its original shape when subjected to corresponding stimuli. Due to the unique properties of shape memory effect, pseudoelasticity, corrosion resistance, and biocompatibility, nickel-titanium (NiTi) SMAs have been vastly applied in the medical and non-medical fields [2,3]. The demand for complex or customized NiTi SMAs structures has facilitated the development of various fabrication technologies, including casting (vacuum arc re-melting, vacuum induction melting, etc.) and powder metallurgy (hot isostatic pressing, metal injection molding, spark plasma sintering, etc.) [4-6]. These existing methods, however, still have some critical limitations that affect their functional performance, including high impurity pickup, material segregation, and restriction on the geometrical complexity. Additive manufacturing (AM) technology through layer by layer deposition can directly manufacture complex three-dimensional solid parts with low or minimum defects and freedom to design complex part geometry with internal features that are otherwise difficult to achieve for casting and powder metallurgy. It offers an ideal solution for the production of NiTi shape memory alloys [7].

Current research on additive manufacturing of NiTi shape memory alloys has been mainly limited to experimental studies. Saedi et al. [8] studied the effects of laser power and scanning speed variation on the microstructure evolution, transformation temperatures, and shape memory response. Their results demonstrated that the sample manufactured with low laser power had significantly higher strain recovery and lower mechanical hysteresis compared to the sample fabricated with high laser power. Ma et al. [9] utilized the controlled hatch distance within the same part to achieve a multi-stage transformation. Through the location-dependent control of thermal history, they created a functional structure with multi-stage special responses subject to different temperature stimuli. Sasan et al. [10] demonstrated that the parameters with the same energy density range but composed of high laser parameters (high laser power adjusted to high scanning speed) and low laser parameters (low laser power adjusted to low scanning speed) would result in different shape memory responses. The lower cooling rate resulted in higher martensite to austenite transformation temperature. Wang et al. [11] fabricated a layer-structured NiTi sample with two alternant sets of process parameters. A much wider transformation temperature range (~80 K) with internal friction tan $\delta$ > 0.04 than that of conventional material (only 25 K) was achieved. The layer-structured sample also shown a weak temperature dependence of storage modulus when the temperature cooled from 373 K to 240 K. The above researches focused on the relationship between the process parameters and the characteristic temperature of NiTi shape memory alloy. The attempted efforts have suggested that high-quality NiTi shape memory parts can be additively manufactured through suitable process parameters. However, a further systematic study and simulation efforts to establish the relationship between the process parameters and final mechanical properties for NiTi shape memory alloy (e.g. a robust model to predict the geometry, thermal history, microstructure evolution, secondary phase precipitation, and characteristic temperature, etc.) are still lacking.

The additive manufacturing process involves convection, conduction, surface radiation, and material phase change, which makes it very hard to predict the final mechanical properties through simple experimental calibration. Therefore, numerical simulation has been widely adopted to study the physical process (e.g., element distribution, geometry formation, thermal gradient, etc.) and to predict the defects formation (e.g., cracking, dislocation, weak metallurgical bonding, inter/intralayer porosity, nickel evaporation, precipitation, etc.) in the deposition process [12-16]. Tian et al. [17] established a multi-physics model through predefined melt pool shape to investigate the influence of process parameters on the melt pool geometry and dilution ratio. A linear relationship between the relative energy-mass ratio and the dilution ratio was found in their results. Song et al. [18] developed a numerical model to predict the melt-pool surface curvature and solidified bead dimensions. The laser power attenuation caused by powder flux and the temperature of powder before it falling into the molten pool were considered in their model. The result showed a maximum error of 10% between simulation and experimental data under a wide range of processing parameters. Shi et al. [19] adopted a transient thermal dynamic model based on the curved substrate to investigate the molten pool behavior in the process of coating the inner surface of the steel tube. The results indicated that the length of the molten pool was restricted by the curved surface of the substrate, yielding a limited wetting angle caused by the weakening of gravity. Shao et al. [20] utilized a three-dimensional model considering heat transfer, phase change, and Marangoni flow to illustrate the grain structure evolution. The simulated cooling rate (temperature gradient $G$ x solidification rate $R$) was consistent with the grain structure distribution in the sample cross-section. Fallah et al. [21] employed a quantitative phase-field (PF) model coupled with the heat transfer finite element model to predicate the local dendrite growth along with the solid/liquid (S/L) interface. The simulated primary dendrite arm spacing value under the directional solidification condition matched well with the experimental results. Moreover, in terms of predicting minimum spacing value, the proposed PF model and the analytical model of McCartney and Hunt [22] obtained the same results. Based on the solute transport and mass conservation theory, He et al. [23] simulated the evolution of concentration distribution of carbon and chromium elements in the molten pool. Through this self-consistent model, they proved that the final concentration distribution of elements was primarily determined by the melting stage. Chen et al. [24] proposed an improved multi-physics FE model considering the shielding gas pressure and powder temperature-rise to predict 3D geometry and solidification behavior in the Fe-based coating process. The results revealed that the crystal structures were significantly influenced by the temperature gradient and undercooling at the solidification interface, while the grain size is mainly determined by the solidification rate.

Up to now, most DED simulation models have two problems. One is the adoption of pre-defined cladding shape as input to study the temperature field or fluid velocity, which limits the model predictability. The other is the laser-powder interaction that results in complex attenuation or absorption coefficients for energy and mass input. Current methods to deal with these coefficients are mainly through constant assumption or numerical analysis based on the concentration of powder distribution. However, the concentration of powder distribution is hard to measure. The existing formula for the powder concentration distribution does not consider the influence of protective gas on the powder distribution and the influence of decentration between power and powder flux on power attenuation, which are also very important. On the other hand, some of the experimental measurements for the previous coefficients were conducted, but the laser power absorptivity is usually separately calculated through the Hagen-Rubens relationship [25], where the temperature-varying resistivity of mixture materials in the molten pool is difficult to calculate in the process. Therefore, an alternative method to deal with the previous problems is desired. For the molten pool evolution, the growth of the molten pool surface can be realized through the Arbitrary Lagrangian-Eulerian (ALE) moving mesh method and its growth velocity will be determined by the added powder mass in the molten pool. The interface of solid and liquid can be distinguished through the viscosity variation. For the laser-powder interaction, because it is difficult to distinguish whether the energy entering the molten pool is directly absorbed from the laser power or brought in by the powder, an effective absorption coefficient can be utilized to calculate the total energy input in the molten pool, as more closely to the actual situation.

In this paper, a transient three-dimensional model including heat transfer, phase change, and fluid flow is proposed for the direct energy deposition of NiTi shape memory alloy. The free surface with Arbitrary Lagrangian-Eulerian (ALE) moving mesh method is employed to track the molten pool growth. Moreover, the model adopts the laser power effective absorption coefficient measured through the calorimetric method and powder absorption coefficient determined from the weight method to deal with the laser-powder interaction and mass adding, as to provide a fast and reliable method for the model simulation. Thus, this paper provides novelty as compared to other models for DED process simulation. The cladding geometry, molten pool depth, and deposition rate were simulated under different process parameters and compared with experimental observations. The fluid velocity was investigated to predict the element distribution. The temperature gradient *G*, solidification rate *R*, as well as shape factor *G/R*, were calculated. Its influence on the microstructure morphology distribution was discussed. This model allows us to optimize the deposition process parameters and can give a better understanding of grain structure evolution in the melt pool solidification process.

## 2. Numerical modeling

Fig. 1 shows the schematics of directed energy deposition. A laser beam with high energy density scans across the substrate surface and creates a melting pool on its optical focus. Simultaneously, the powder stocks are injected into the molten pool by a carrier inert gas through a coaxial powder nozzle. The inert gas is also used to protect the molten pool from oxidation from the surrounding environment. When the laser spot moves away, the molten pool begins to solidify to form the cladding track. In the whole deposition process, multi-physics phenomena are involved including heat transfer in solid and liquid (e.g. convection, conduction, and surface radiation) and fluid flow in the molten pool.

Therefore, to simplify the simulation model, the following assumptions are made:

(1) The fluid flow in the molten pool is assumed to be Newtonian, laminar, and incompressible;
(2) The input laser power is assumed to have a Gaussian distribution and is added to the molten pool surface. The beam radius and profile are kept constant along the z-direction. The effective absorption coefficient remained constant throughout the whole process;
(3) There is no diffusion transport in the solid and liquid phases and vaporization is ignored;
(4) The coaxial powder flow is assumed to have a Quasi-Gaussian distribution. The shape of powder particles is spherical. Powders falling into the region of the molten pool are melted immediately while the momentum is zero.
(5) Laser and powder stream are always in focus on the substrate surface. The vertical deviation of the focal spot is ignored.

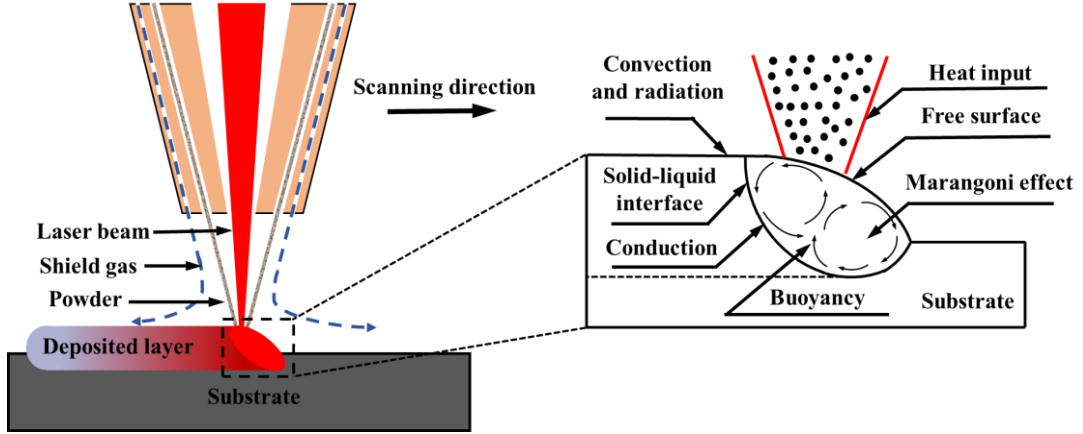

Fig. 1. Schematic of directed energy deposition and its involved multi-physics. phenomena.

### 2.1 Governing equations

A transient three-dimensional finite element (FE) code with COMSOL Multiphysics 5.4 was developed to simulate the directed energy deposition process. The schematic of the FE model was shown in Fig. 2. The laser beam and powder jet moved along the predefined trajectory while the substrate was left stationary. The substrate in the model had dimensions of 50 mm (L) x 25 mm (W) x 4 mm (H) as those in the experiments. The computational domain was divided into two domains: solid domain and fluid domain, wherein the whole process was governed by the energy conservation Eq. (1), momentum conservation Eq. (2), and mass conservation Eq. (3) [26,27].

$$\rho c_p \left( \frac{\partial (T)}{\partial t} + \boldsymbol{u} \cdot \nabla T \right) = \nabla \cdot (k \nabla T) + \dot{Q}(x, y, z) \quad (1)$$

$$\rho \left[ \frac{\partial \boldsymbol{u}}{\partial t} + (\boldsymbol{u} \cdot \nabla) \boldsymbol{u} \right] = \nabla \cdot [-p \boldsymbol{I} + \mu (\nabla \boldsymbol{u} + \nabla \boldsymbol{u}^T)] + \boldsymbol{F} \quad (2)$$

$$\frac{\partial \rho}{\partial t} + \nabla \cdot (\rho \boldsymbol{u}) = \dot{S}_{powder} \qquad (3)$$

where $\boldsymbol{u}$ is the velocity (m/s); $\dot{Q}(x,y,z)$ is the external energy source (W); $p$ is the pressure (Pa); $\dot{S}_{powder}$ is the external mass source term caused by powder addition (kg/s); and $\rho$, $c_p$, $k$, $\mu$ are the equivalent density (kg/m³), heat capacity (J/(kg*K)), thermal conductivity (W/(m*K)), and viscosity (Pa*s), respectively, which are related to temperature and z-position and defined as follows [14,28]:

$$\rho = step\left(\frac{z_0-z}{\Delta z}\right) * \left[\rho_1 * \left(1 - step\left(\frac{T-T_{m1}}{T_{l1}-T_{s1}}\right)\right) + \rho_2 * step\left(\frac{T-T_{m1}}{T_{l1}-T_{s1}}\right)\right] + step\left(\frac{z-z_0}{\Delta z}\right) * \rho_2 \qquad (4)$$

$$k = step\left(\frac{z_0-z}{\Delta z}\right) * \left[k_1 * \left(1 - step\left(\frac{T-T_{m1}}{T_{l1}-T_{s1}}\right)\right) + k_2 * step\left(\frac{T-T_{m1}}{T_{l1}-T_{s1}}\right)\right] + step\left(\frac{z-z_0}{\Delta z}\right) * k_2 \qquad (5)$$

$$c_p = step\left(\frac{z_0-z}{\Delta z}\right) * \left[c_1 * \left(1 - step\left(\frac{T-T_{m1}}{T_{l1}-T_{s1}}\right)\right) + c_2 * step\left(\frac{T-T_{m1}}{T_{l1}-T_{s1}}\right) + L_{m1}\frac{Rect(T)}{T_{l1}-T_{s1}}\right]$$
$$+ step\left(\frac{z-z_0}{\Delta z}\right) * \left(c_2 + L_{m2}\frac{Rect(T)}{T_{l2}-T_{s2}}\right) \qquad (6)$$

$$\mu = step\left(\frac{z_0-z}{\Delta z}\right) * \left[\mu_s * \left(1 - step\left(\frac{T-T_{m1}}{T_{l1}-T_{s1}}\right)\right) + \mu_{l1} * step\left(\frac{T-T_{m1}}{T_{l1}-T_{s1}}\right)\right]$$
$$+ step\left(\frac{z-z_0}{\Delta z}\right) * \left[\mu_s * \left(1 - step\left(\frac{T-T_{m2}}{T_{l2}-T_{s2}}\right)\right) + \mu_{l2} * step\left(\frac{T-T_{m2}}{T_{l2}-T_{s2}}\right)\right] \qquad (7)$$

$$step(x) = \begin{cases} 0 & x < -0.5 \\ smooth\ transition\ function\ from\ 0\ to\ 1 & -0.5 \leq x \leq 0.5 \\ 1 & x > 0.5 \end{cases} \qquad (8)$$

Here, $z_0$ is the initial plane of the substrate (mm); $\Delta z$ is a small constant associated with the cladding height (mm); subscripts *1* and *2* represent the substrate and powder materials respectively; *s* and *l* denote the solid and liquid phases, respectively; $T_m$(K) is the melting temperature, which is assumed to be the average value of solidus temperature $T_s$ (K) and liquidus temperature $T_l$ (K); $L_m$ is the latent heat enthalpy of fusion (J/kg); *Rect(T)* is a rectangle function that is equal to 1 within the range of $T_{s1,2}$ (K) to $T_{l1,2}$ (K) and 0 outside the range; $step(x)$ is a smooth step function. The materials of substrate and powder are distinguished by the z-potion and molten temperature. Above the initial surface $z_0$ the material is the powder and below the surface $z_0$, the material is the mixture of powder and substrate, which is determined by the temperature.

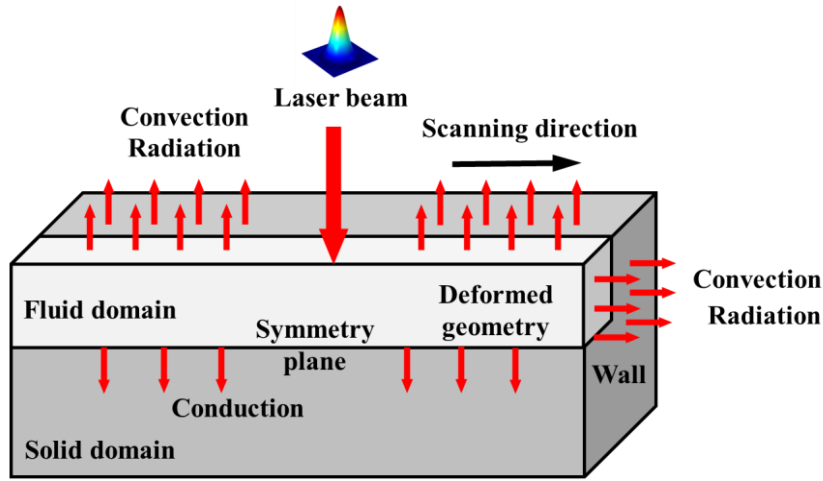

Fig. 2. Schematic of FE model geometry.

The volume force vector **F** represents the buoyancy and gravity force, which is given as follows [29]:

$$\boldsymbol{F} = \rho_0(1 - \beta(T - T_0))\boldsymbol{g} \tag{9}$$

in which $\rho_0$ is the density at the reference temperature (kg/m³); $\beta$ is the thermal expansion coefficient (1/K); $T$ (K) and $T_0$ (K) are the transient temperature and reference temperature, respectively.

### 2.2 Initial and boundary conditions

The substrate is mounted on the machine platform. The platform has a large bulk volume and its surrounding environment is infilled by inert argon gas. Hence, the bottom surface of the substrate is assumed to be thermally insulated and be kept at room temperature 293.15 K. The top surface is exposed to external heat source input, heat convection, and surface to ambient radiation. The thermal boundary condition is described by the following equation [30]:

$$k(T)(\nabla T \cdot \boldsymbol{n}) = Q(x,y) - h(T - T_0) - \varepsilon\sigma_b(T^4 - T_0^4) \tag{10}$$

in which $h$ is the convection coefficient and the value of 25 W/(m² *K) is adopted in our simulation [31]; $\varepsilon$ is surface emissivity; $\sigma_b$ is the Stefan-Boltzmann constant [W/(m²*K⁴)]; and $T_0$ is the room temperature at 293.15 K.

The source term of $Q(x,y)$ is very complex due to the interaction of laser power and powder flow (W/m²). When the laser beam passes through the powder flow, part of the energy will be absorbed and reflected or scattered by powder particles, while the remaining energy portion then passes through the powder flux to illuminate the substrate surface and create the molten pool. At the same time, the scattered powders will absorb energy to heat themselves. The energy comes from the power beam flux and the radiation reflected from the substrate. After the powder arrives at the substrate, a portion falls into the molten pool while the remaining are assumed to rebound back to the surrounding environment. Therefore, the heat source $Q(x,y)$ acted on the pool surface can be expressed by:

$$Q(x,y) = Q_{laser} + Q_{powder} \tag{11}$$

$$Q_{laser} = I(x,y) * \alpha_{laser} * (1 - x_{atten}) \tag{12}$$

$$Q_{powder} = I(x,y) * \{\alpha_{powder} * x_{atten} + \alpha_{powder} * (1 - \alpha_{laser}) * (1 - x_{atten})\} * \eta_m \tag{13}$$

$$I(x,y) = \frac{2P_l}{\pi * r_b^2} \exp\left(-\frac{2r^2}{r_b^2}\right) \tag{14}$$

where $I(x,y)$ is the laser power intensity assumed to have a Gaussian distribution (W/m²); $\alpha_{laser}$ is the laser absorption coefficient; $x_{atten}$ is the powder attenuation coefficient caused by powder particle absorption or reflection; $\alpha_{powder}$ is the powder particle absorption coefficient for the energy from laser illumination, reflection or scattering from neighboring particles and substrate; $\eta_m$ is the powder absorption coefficient; $P_l$ is the input laser power (W); $r_b$ is the effective laser energy beam radius (m); $r$ is the distance between the heating source center and the current computational cell (m).

The laser power effective absorption coefficient is defined by:

$$\eta = \frac{Q_{laser} + Q_{powder}}{P_l} \tag{15}$$

The around side surfaces are only subjected to heat convection and surface to ambient radiation. Their relation is given by [32]

$$k(T)(\nabla T \cdot \boldsymbol{n}) = -h(T - T_0) - \varepsilon\sigma_b(T^4 - T_0^4) \tag{16}$$

When the melt pool is formed, its surface is subject to capillary force and thermocapillary force. The capillary force (surface tension) acts in the normal direction while the thermocapillary force (Marangoni effect) exerts on the tangential direction. The force balance equation on the surface of the melt pool is given as follows[18] :

$$\mathbf{T} \cdot \boldsymbol{n} = \sigma(\nabla_t \cdot \boldsymbol{n})\boldsymbol{n} - \frac{\partial\sigma}{\partial T}\nabla_t T \cdot \boldsymbol{t} \tag{17}$$

where **T** is the stress tensor on the open boundary; $\boldsymbol{n}$ is the unit normal vector of the melt pool surface; σ is the surface tension (N/m); $\nabla_t \cdot \boldsymbol{n}$ is the curvature of the surface profile; $\frac{\partial\sigma}{\partial T}$ is the thermocapillary coefficient (which is the temperature derivative of the surface tension) (N/(m*K)); $\nabla_t T$ is the temperature gradient along the tangential direction of the melt pool surface; $\boldsymbol{t}$ is the unit tangential vector of the local pool surface.

To simulate the deformation of the melt pool surface, Arbitrary Lagrangian-Eulerian (ALE) moving mesh method is adopted. This method allows the boundary nodes of the molten pool surface to vary along with the powder feed rate and fluid dynamics. The moving mesh boundary condition is given as follows [33]:

$$\boldsymbol{u}_{mesh} \cdot \boldsymbol{n} = \boldsymbol{u} \cdot \boldsymbol{n} - \frac{2m_f\eta_m}{\rho_m\pi r_p^2}exp\left(\frac{-2r^2}{r_p^2}\right) * step(\frac{T-T_{m1}}{T_{l1}-T_{s1}}) \cdot \boldsymbol{n} \tag{18}$$

where $\boldsymbol{u}_{mesh}$ is the mesh velocity of the melt pool surface (m/s); $\boldsymbol{u}$ is the material velocity computed from the momentum conservation Equation (2) (m/s); $m_f$ is the powder feed rate (kg/s); $\eta_m$ is the powder absorption coefficient; $r_p$ is the powder beam radius (m); These three parameters were determined through experiments, $\rho_m$ is the powder density (kg/m³).

To save calculation time, only half of the substrate domain is considered in the model, therefore the symmetry boundary conditions for heat transfer in solid and liquid and fluid flow in the molten pool are given as follows [29]:

$$\underline{\boldsymbol{n}} \cdot (\lambda\nabla T) = 0 \tag{19}$$

$$\boldsymbol{u} \cdot \underline{\boldsymbol{n}} = 0 \tag{20}$$

$$\{\mu[\nabla\boldsymbol{u} + (\nabla\boldsymbol{u})^T]\}\underline{\boldsymbol{n}} - \left(\left(\{\mu[\nabla\boldsymbol{u} + (\nabla\boldsymbol{u})^T]\}\underline{\boldsymbol{n}}\right) \cdot \underline{\boldsymbol{n}}\right)\underline{\boldsymbol{n}} = 0 \tag{21}$$

where $\underline{\boldsymbol{n}}$ is the normal vector defined on the symmetry plane; These boundary conditions set the heat transfer and fluid flow equal to zero at the symmetry plane.

**2.3 Parameters and material properties**

In the simulation model, the material of the substrate was set to pure titanium with the weight % composition of Fe (≤0.2), C (≤0.08), O (≤0.15), N (≤0.03), H (≤0.013), and Ti (balance). While the material of powder was given by pre-alloyed NiTi (Ni 50.93 at %) powder with the weight % composition of Ni (56), Fe (0.0074), C (0.006), O (0.0607), N (0.0018), and Ti (balance). The pre-alloyed powder has a similar composition with the Nitinol and other powders (e.g., $Ni_{50.99}Ti_{49.91}$ [34], $Ni_{50.8}Ti_{49.2}$ [35], etc.) reported in the literature, which is typically composed of about 50 to 51% nickel by atomic percent (55-56% weight percent). The characteristic transformation temperatures of the adopted powder are shown in Table 1. The appearance of a two-step transformation is attributed to the existence of the $Ni_4Ti_3$ phase, which will result in the separation of the R phase from the B2↔B19' transformation [36,37].

**Table 1**
The characteristic transformation temperature of the adopted powder.

| Sample | Martensitic transformation temperature | | | | Austenitic transformation temperature | | | |
|---|---|---|---|---|---|---|---|---|
| | $R_s$ | $R_f$ | $M_s$ | $M_f$ | $R_s$ | $R_f$ | $A_s$ | $A_f$ |
| NiTi powder | 4.5°C | -9.8°C | -16.3°C | -60.3°C | -43.5°C | -22.4°C | -18.7°C | 22.9°C |

The powder absorption coefficient $\eta_m$ was determined through the weight method. For each deposition parameter, three single tracks with 15 mm length were deposited on the substrate to calculate the single powder absorption coefficient. Three tests were repeated to finally calculate the powder absorption coefficient $\eta_m$. Through this method, the defocus between powder flux and laser beam as well as the burning loss can be effectively compensated in the simulation model. The laser power effective absorption coefficient $\eta$ was measured through the calorimetric method, as described by Lia et al. [38], Wirth et al. [39], and Trapp et al. [40]. For each deposition parameter, a single track with 10 mm length was fabricated on the substrate and the temperature increment of the substrate with clad powder was measured through the thermocouple to calculate the total absorbed energy. Three tests were repeated to finally calculate the laser power effective absorption coefficient $\eta$. The calculated powder absorption coefficient $\eta_m$ and the laser power effective absorption coefficient $\eta$ were shown in Fig. 3 and Fig. 4. As shown in Fig. 3, the powder absorption coefficient increases along with the increment of laser power. This phenomenon that is attributed to the fact that a larger laser power will create a larger molten pool under the powder projection area, absorbing more powder particles. In Fig. 4, it is found that the effective absorption coefficient is relatively stable when the powder feed rate is less than 0.96 g/min but decreases rapidly along with the increment of laser power after reaching 0.96 g/min. This situation is related to the energy loss caused by the powder flux, which is proportional to the powder bundle density and input laser power. Other thermophysical properties of materials were given in Table 2 and Fig. 5. Due to the lack of relevant data in the high-temperature range for NiTi, the temperature-dependent material properties in Fig. 5 were evaluated based on the calculation of phase diagrams through JMatPro software, which was also used by Chen et al. [24] in their model to calculate the temperature-dependent properties of iron (Fe)-based powder and 45 # stainless steel substrate.

For model accuracy and convergence, the maximum mesh size of 20 μm was assigned in the fluid domain while the default normal mesh was given to the solid domain. The governing equations were discretized using P2+P1 default settings for laminar flow and linear temperature for heat transfer. The solution was performed with the PARDISO solver and Arbitrary Lagrangian-Eulerian (ALE) moving mesh was smoothed by the Laplace method. The variables such as temperature, pressure, velocity, etc. were calculated in a fully coupled manner. The time step value guaranteed the maximum displacement of the free surface is less than the minimum grid spacing for each step. The smooth step function was applied to the transition from solid to liquid for all thermophysical properties. High dynamic viscosity of

100 Pa*s was assigned to the solid phase in the fluid domain and solid domain.

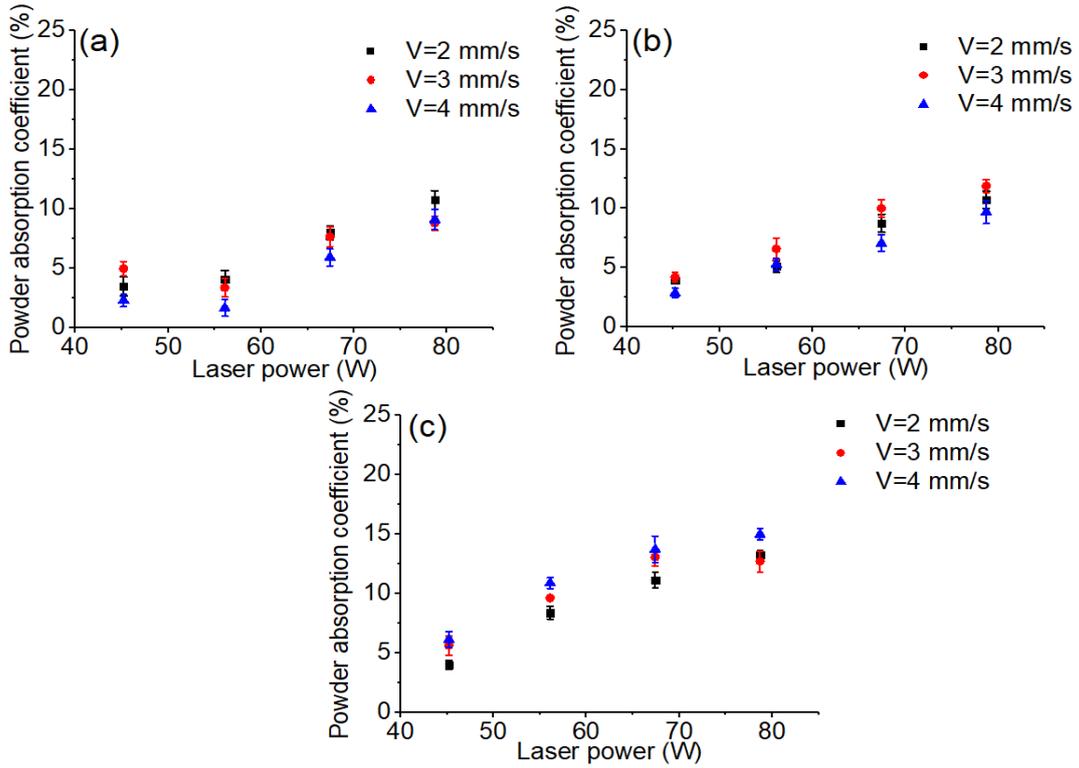

Fig. 3. The powder absorption coefficients vary along with laser power and scan speed at different powder feed rates: (a) 0.69 g/min, (b) 0.83 g/min, and (c) 0.96 g/min.

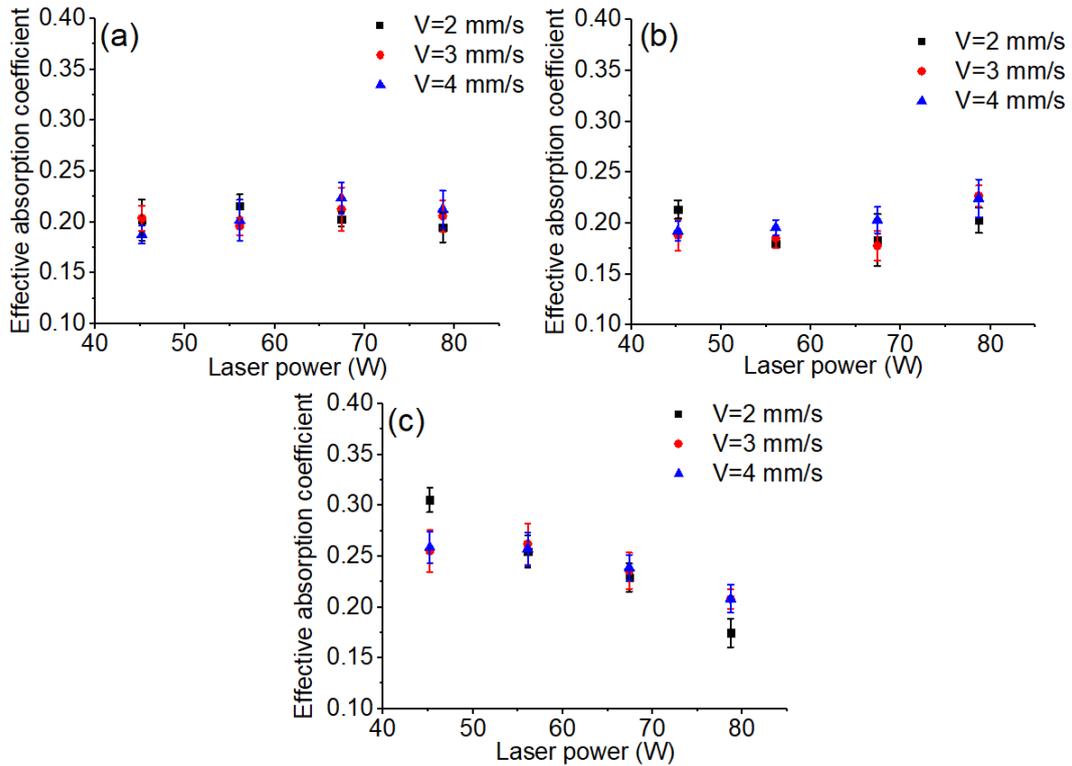

Fig. 4. The laser power effective absorption coefficients vary along with laser power and scan speed at different powder feed rates: (a) 0.69 g/min, (b) 0.83 g/min, and (c) 0.96 g/min.

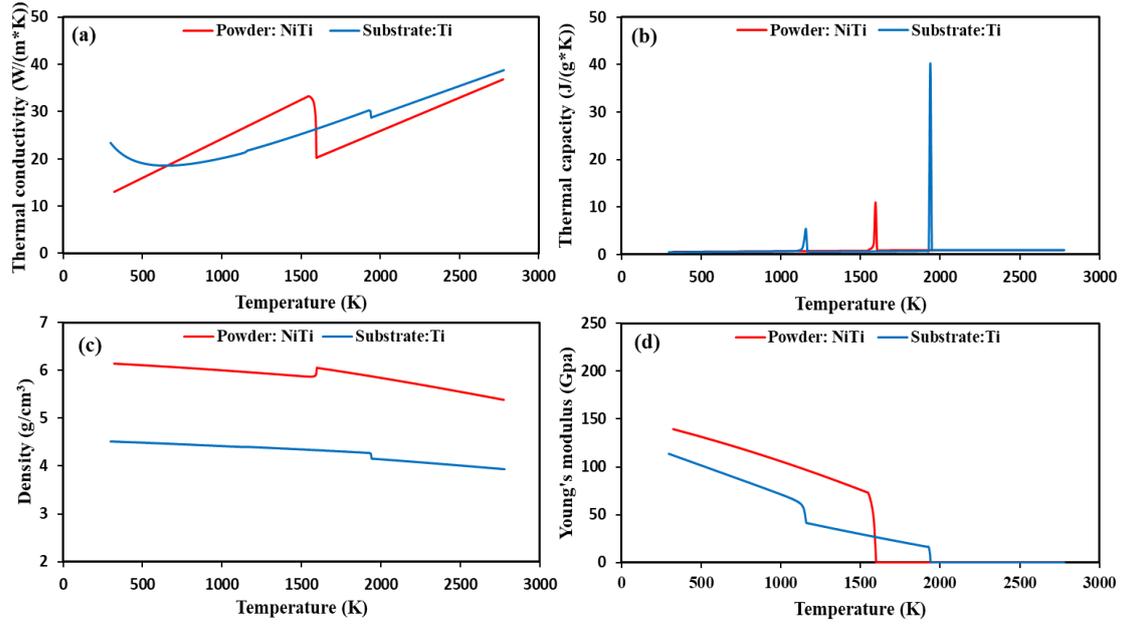

Fig. 5. Thermal-dependent material properties: (a) thermal conductivity, (b) thermal capacity, (c) density, and (d) Young's modulus.

**Table 2**
Material properties for powder and substrate.

| Parameter | Pure titanium [41] | NiTi powder [42-44] |
|---|---|---|
| Density, $\rho_1, \rho_2$ | T-dep, 3.9-4.5 g/cm$^3$ | T-dep, 5.4-6.45 g/cm$^3$ |
| Thermal conductivity, $k_1, k_2$ | T-dep, 17.1-28 W/(m*K) | T-dep, 13-36 W/(m*K) |
| Specific heat capacity, $c_1, c_2$ | T-dep, 520-967 J/(kg*K) | T-dep, 470-840 J/(kg*K) |
| Thermal expansion coefficient, $\beta$ | T-dep, 8.4-10.2 e-6 1/K | 11e-6, 1/K |
| Solidus temperature, $T_{s1}, T_{s2}$ | 1931 K | 1553 K |
| Liquidus temperature, $T_{l1}, T_{l2}$ | 1951 K | 1583 K |
| Latent heat, $L_{m1}, L_{m2}$ | 295.6 kJ/kg | 24.2 kJ/kg |
| Liquid viscosity, $\mu_{l1}, \mu_{l2}$ | 4.42-0.00667*(T-1943) mPa*s | 5.74 mPa*s |
| Surface tension, $\sigma$ | 1.557-0.00016*(T-$T_{l1}$) N/m | 1.78-0.00025*(T-$T_{l2}$) N/m |

## 3. Experimental setup

To validate the simulation model, a series of single tracks were deposited under different process parameters on the pure titanium substrate of 50 x 25 x 4 mm$^3$. The process parameters were given in Table 3. The deposition process was conducted on a 6-axis parallel platform (PI H840.D11, Physik Instrument) equipped with a multi-mode continuous-wave laser (wavelength: 1070 nm, YLR-500-MM-AC-Y14, IPG Photonics) of the maximum output power of 500 W. The laser was focused on the substrate surface through an optical delivery system (KUKA Laser Industry) with a working distance of 12 mm. Meanwhile, the powder stock was fed by a disk-type feed system (GPV PF2/2, GTV Thermal Spray). In the deposition process, a 3 L/min argon gas was adopted as a carriage to transport powder material and a 10 L/min argon gas was used as the protective gas to prevent melt pool oxidation. The experimental setup was developed by our group as shown in Fig. 6 [45].

**Table 3**

Data used in simulation and experiment.

| Processing parameters | Value |
| --- | --- |
| Laser Power $P_l$ (W) | 45.2, 56.1, 67.4, 78.7 |
| Powder feed rate $v_{powder}$ (g/min) | 0.69, 0.83, 0.96 |
| Scan speed $v_p$ (mm/s) | 2, 3, 4 |
| Laser beam radius $r_b$ (μm) | 200 |
| Powder beam radius $r_p$ (μm) | 400 |
| Laser wavelength $\lambda$ (nm) | 1070 |

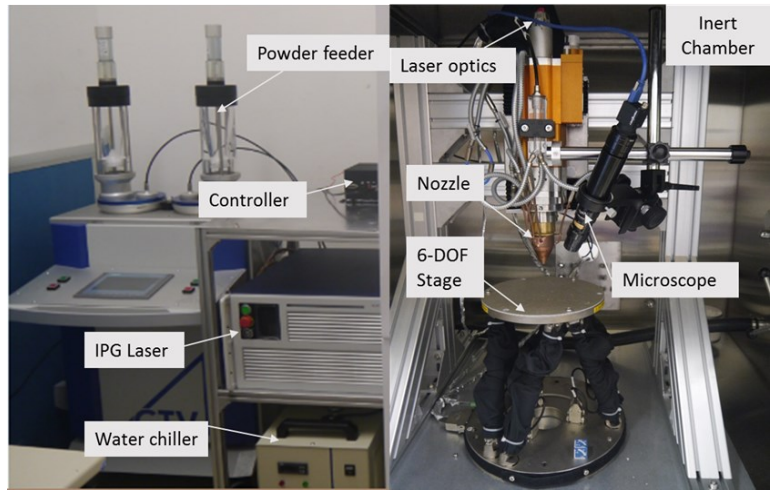

Fig. 6. Direct energy deposition system [45].

The samples were cut using electrical discharge machining (EDM) and polished through SiC abrasive paper with grit sizes of 180, 400, 600, 800, 1200, then polished using 0.5 μm diamond suspensions on synthetic cloths. The etching solution was a 1:5:40 volume mixture of HF, $HNO_3$, and $H_2O$, respectively. The cross-sections of the melt pool were inspected through optical microscopy (RH-2000, Hirox), while microstructures of the etched cross-section were captured by scan electron microscope (JEOL JSM 6400).

## 4. Results and discussions

### 4.1 Clad geometry

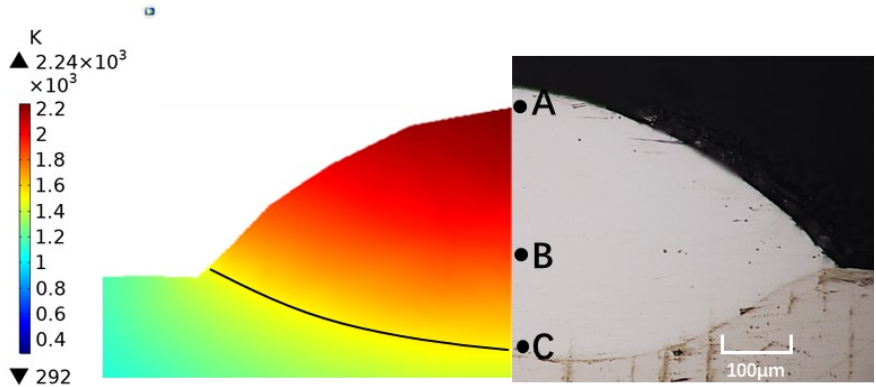

Fig. 7. Comparison of cross-sectional profile of experimental and simulation results for coated powder cladding under laser power 56 W, powder feed rate 0.83 g/min, and scan speed 2 mm/s.

Fig. 7 shows the simulated and experimental molten pool morphology in the cross-section perpendicular to scan direction with deposition parameters of laser power 56 W, powder feed rate 0.83 g/min, and scan speed of 2 mm/s. The black-colored line in Fig. 7 corresponds to the liquid-solidus interface. The height of the cladding layer was 120 μm in the simulation and 135 μm in the experiment. The width of the cladding layer was 422 μm in the simulation and 456 μm in the experiment. The depth of the cladding layer was 57 μm and 63 μm in the simulation and experiment respectively. The deviation of height, width, and depth was 11.1%, 7.5%, and 9.5%.

To verify the robustness of the numerical model, a series of single-tracks were deposited with different process parameter combinations. The cladding geometry and melt pool depth were plotted in Fig. 8. The results presented in Fig. 8(a) were obtained under a constant scan speed of 2 mm/s while the curves in Fig. 8(b) were derived under a constant laser power of 45.2 W. The powder feed rate for both cases was chosen as 0.83 g/min. Meanwhile, the results in Fig. 8(c) were conducted under a constant laser power of 45.2 W and a scan speed of 2 mm/s.

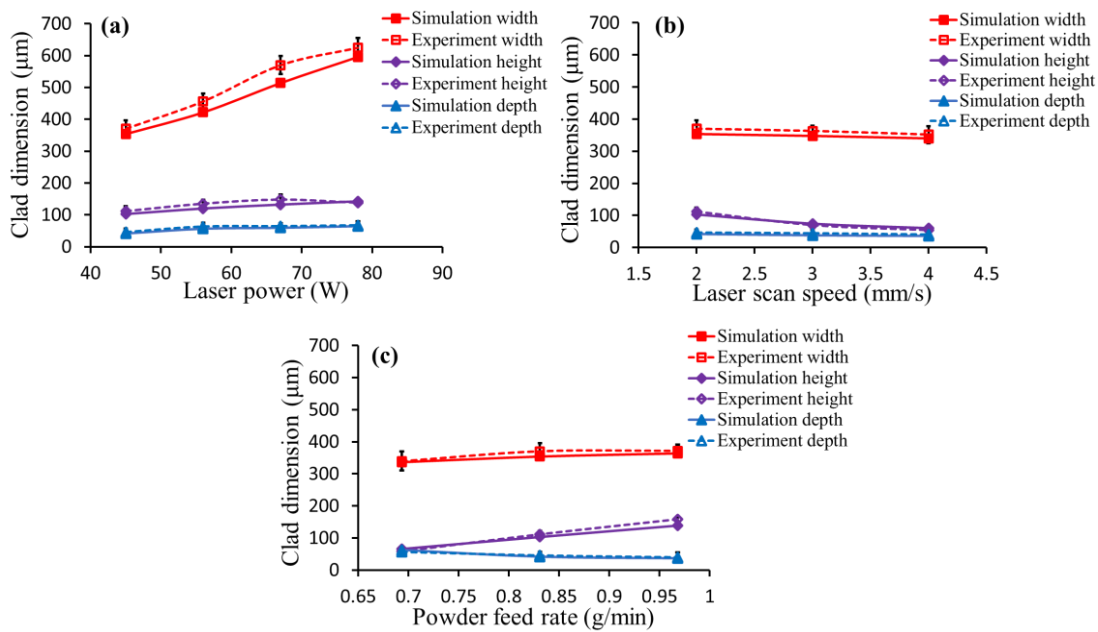

Fig. 8. Simulation and experimental results of the deposited single tracks under: (a) varying laser power, (b) varying laser scan speed, and (c) varying powder feed rate.

It can be seen from Fig. 8(a) that when the laser power increases from 45.2 W to 78.7 W, the width increases rapidly while the height and depth have a slight influence. This increase of melt-pool width is primarily caused by the higher laser energy input because a higher energy input will melt more materials on the substrate and create a wider molten pool as well as improve the powder absorption coefficient. Meanwhile, according to the references [41,42], the thermo-capillary gradient of NiTi alloy is negative, which indicates convective melt-pool fluid flows from the center area to the pool edge and will result in a wider and shallower morphology [46,47]. A typical molten pool velocity field in our experiments is shown in Fig. 9. The arrows indicate the flow direction and the arrow length is proportional to the velocity. It is observed that the fluid flows from the center area to the pool edge which broadens the clad width and weakens the increase of height and depth caused by higher laser power. Besides, it is easy to see that the deviation of width between simulated and experimental results in Fig. 8(a) is larger than the deviation in Figs. 8(b) and 8(c). This distinction comes from the powder absorption coefficient change which is more sensitive to the variation of laser power. As shown in Fig. 8(b), the clad height decreases along with the increase of scan speed, while the width and depth of the molten pool have slower reduction. This reduction of clad height is caused by the factors of less powder input per unit area and low molten pool temperature along with the increase of scan speed. Although this increase in scan speed results in less laser energy density per unit area, there is also less powder density per unit area and lower fluid flow convection, which makes the width and depth response insensitive to the variation of scan speed. From Fig. 8(c) it is found that increasing the powder feed rate can rapidly increase the clad height and reduce the molten pool depth, while slightly increase the molten pool width. When there is enough energy input, a higher powder feed rate means that material stock will fall into the molten pool. The absorbed laser energy then will be mainly used to melt falling material stock, which impedes the further temperature rise in the molten pool as well as laser penetration on the substrate. Therefore, the height sharply increases and depth rapidly decreases. The width has no distinct increase due to the weakening fluid flow convection caused by the low-temperature distribution.

Within the scope of process parameters studied, the influence of laser power, scan speed, and powder feed rate on the clad geometry can be summarized as follows. The clad width is mainly determined by the laser power and slightly affected by the scan speed and powder feed rate. The clad height is greatly influenced by the ratio of powder feed rate to scan speed and mildly increased along with the increasing laser power. The clad depth is inclined to increase with the growing laser power and decrease with the increase of scan speed and powder feed rate. The maximum deviation of simulated and experimental results for width, height, and depth are 9.8%, 13%, and 13.6%, respectively.

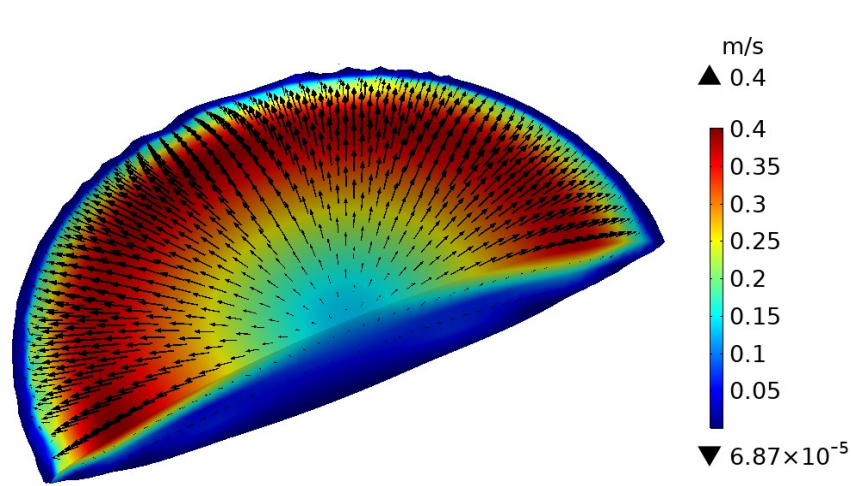

Fig. 9 Molten pool velocity field under process parameter: laser power 56W, scan speed 2 mm/s, and powder feed rate 0.97 g/min.

**4.2 Deposition rate**

Generally, the depth of the melt pool has no distinct influence on the final part geometry. The deposition rate is a more interesting parameter that can estimate cladding time and cost-efficiency. Fig. 10 is the schematic diagram of the cross-section of the cladding layer. The deposition rate is calculated by the following equation:

$$V_{deposition} = S_{clad} * v_p \qquad (22)$$

where $V_{deposition}$ is the deposition volume per unit time; $S_{clad}$ is the cladding cross-section area and $v_p$ is the scan velocity.

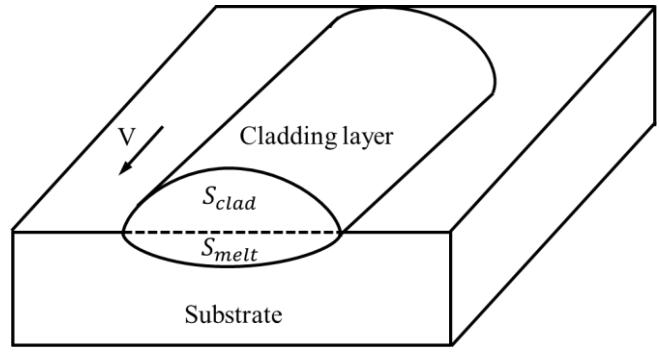

Fig. 10. The schematic of the cladding layer cross section.

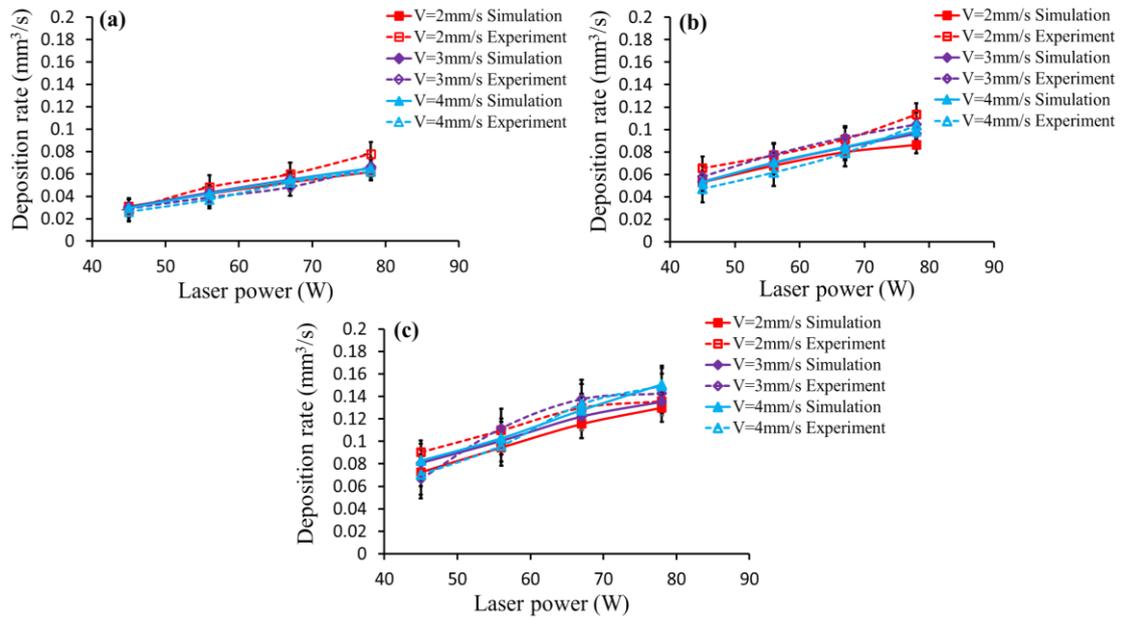

Fig. 11. Simulation and experimental results of the deposited rate under different powder feed rate:
(a) 0.6934 g/min, (b) 0.831 g/min, and (c) 0.968 g/min.

To verify the proposed numerical model, a series of single tracks with different cladding parameters were deposited. The simulated and experimental results of the deposition rate are shown in Fig. 11. From Fig. 11(a) to 11(c), it is observed that the deposition rate sharply increases along with the increasing laser power, regardless of the scan speed variation. This deposition rate improvement can be explained by the wider molten pool and increased powder absorption coefficient caused by the rising laser power. The scan speed has a tiny influence on the deposition rate due to the reciprocal relationship between scan speed and the clad area above the substrate. Besides, under the conditions of the same laser power and scan speed, the increasing powder feed rate can effectively improve the deposition rate. It seems that the higher powder density is beneficial to the convergence of powder flux and provides more opportunity for the powder to fall into the molten pool. Furthermore, within the same powder feed rate, the deviation of simulated and experimental results gradually increases when the laser power increase. This can be attributed to the fact that the intrinsic fluid flow is decided by thermocapillary and temperature gradients which are very sensitive to the laser power variation. Besides, as shown in Fig. 11(c), when the laser power is higher than 67.8 W, the deposition rate measured by the experiment shows a downward inflection point while the value calculated by simulation is still going up. This downward inflection point is caused by the burning of flying powders by high laser power which is not considered in our numerical model.

**4.3 Transport behavior**

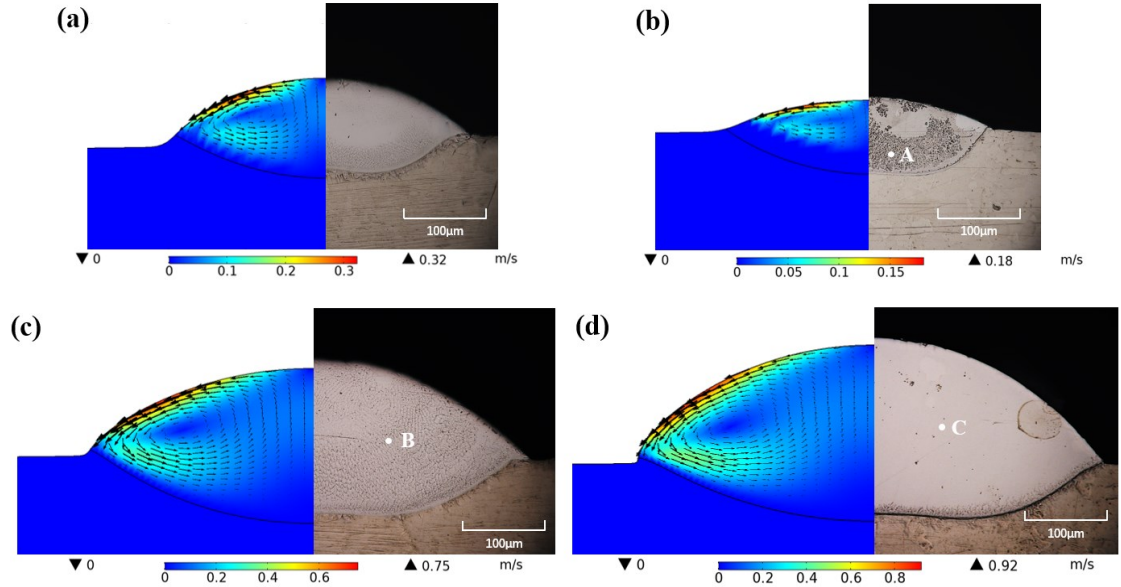

Fig. 12. The velocity fields and element transport behavior: (a) $P_l$=45W, $v_p$=2 mm/s, $v_{powder}$=0.6934 g/min, (b) $P_l$=45W, $v_p$=4 mm/s, $v_{powder}$=0.6934 g/min, (c) $P_l$=67.4W, $v_p$=2 mm/s, $v_{powder}$=0.6934 g/min, (d) $P_l$=67.4W, $v_p$=2 mm/s, $v_{powder}$=0.968 g/min.

The fluid velocity in the molten pool has a significant influence on the distribution of elements in the forming molten pool, which will affect the final microstructure formation. The well-developed velocity field will induce relatively uniform element distribution while a poorly developed velocity field is inclined to result in chemical segregation. To verify this assumption, four different parameter combinations including scan speed, power, and powder variations were chosen to compare in the simulation and experimental results. The cross-sections of the simulated molten pool with velocity fields and corresponding cladding layer after corrosion were shown in Fig. 12. With the presence of a negative thermocapillary coefficient, an outward fluid flow pattern was observed in all four molten pools but the velocity fields present different intensities. From Fig. 12(a) to Fig. 12(b), it is observed that the fluid velocities of both have not fully developed as indicated around the solidification interface, and the fluid velocity further decreases along with the increasing scan speed. This suppressed velocity field will result in different degrees of uneven distribution of elements from substrate Ti and molten powder NiTi as shown in the cross-sections of samples. Due to the smaller velocity field, a more serious microstructure stratification was observed in Fig. 12(b), where the lumpy NiTi phase around the Ti$_2$Ni phase appears in the cladding layer as shown in Fig. 13(a). Compared with Fig. 12(a) and 12(c), it is obvious to see that along with the increasing laser power, a well-developed field with higher fluid velocity is formed, which will further improve the element uniformity. Hence the cross-section of the sample in Fig. 12(c) presents a continuous microstructure distribution without stratification. The zoomed area of point B in Fig. 12(c) is shown in Fig. 13(b). A typical grain morphology with a single-phase indicates the mix of Ti + NiTi → Ti$_2$Ni is fully developed. This sufficient mass exchange in the molten pool is beneficial from the stronger velocity field caused by higher laser power input, promoting the fluid convection through the whole area. As seen from Fig. 12(c) to Fig. 12(d), a higher well-developed fluid velocity field is observed after increasing the powder feeding rate but a smooth cross-section without traces of corrosion is obtained. The enlarged view of point C in Fig. 12 (d) is shown in Fig. 13(c). It is found that a fine NiTi phase and Ti$_2$Ni phase are evenly and alternatively distributed throughout the whole area. This phenomenon is ascribed to the limited Ti melting and redundant NiTi powder adding, inducing the saturation reaction of NiTi + Ti → Ti$_2$Ni. The newly formed distribution shows excellent resistance to the etching solution as indicated by Fig. 12(d).

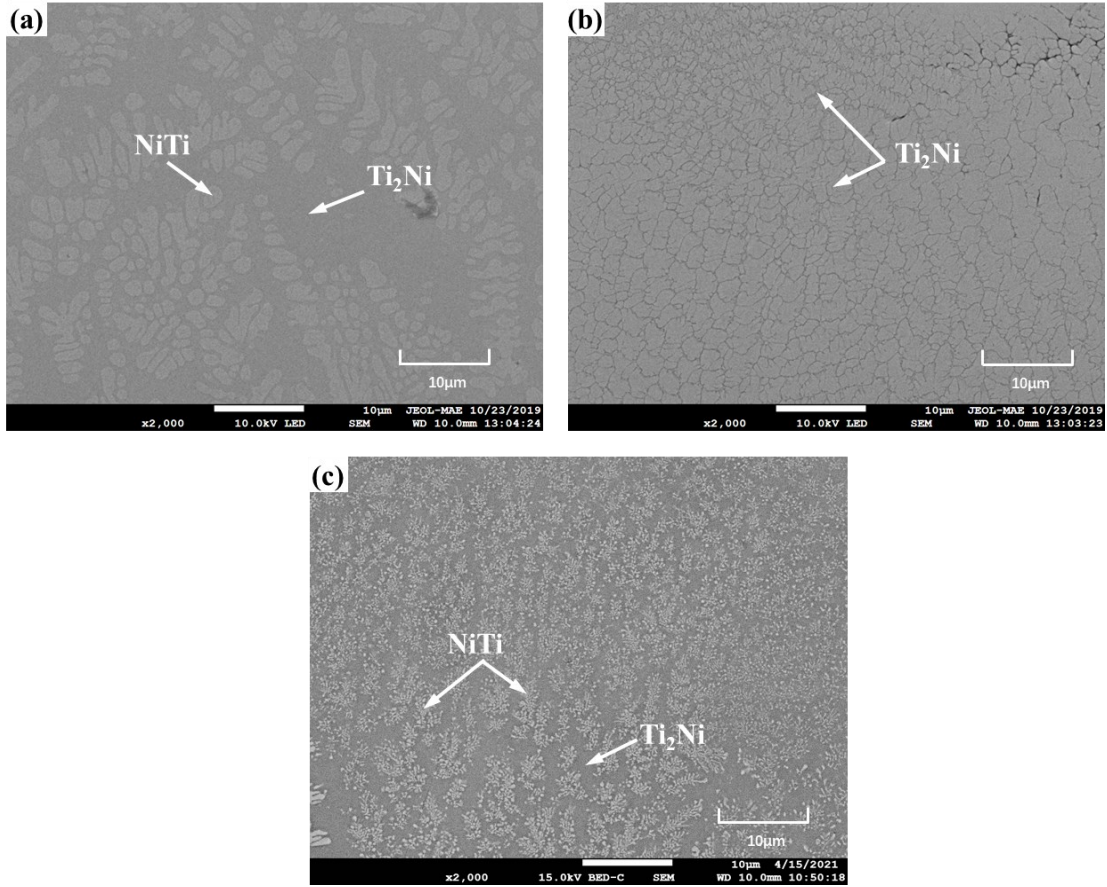

Fig. 13. The phase constituents of formed bead in Fig. 12: (a) point A, (b) point B, and (c) point C.

### 4.4 Solidification characteristics

The temperature gradient normal to the liquid-solid interface **G** and the solidification rate of liquid-solid interface **R** are two important parameters, which determine the solidification microstructures as shown in Fig. 14 [12,33,48]. The product of **G** x **R** determines the grain size, while the quotient of **G** / **R** determines grain morphology. The morphology of the microstructure changes from the planar to cellular to columnar to equiaxed dendritic as the **G** / **R** decreases. The grain size varies from coarser structure to finer structure with the increase of **G** x **R**. The definition of **G** and **R** in the deposition energy deposition method was given as follows:

$$\boldsymbol{G} = \nabla T \cdot \boldsymbol{n} \tag{23}$$

$$\boldsymbol{R} = v_p \cdot \boldsymbol{i} \cdot \boldsymbol{n} \tag{24}$$

where $\boldsymbol{n}$ is the unit normal vector of solidification front and $\boldsymbol{i}$ is the vector of scan direction.

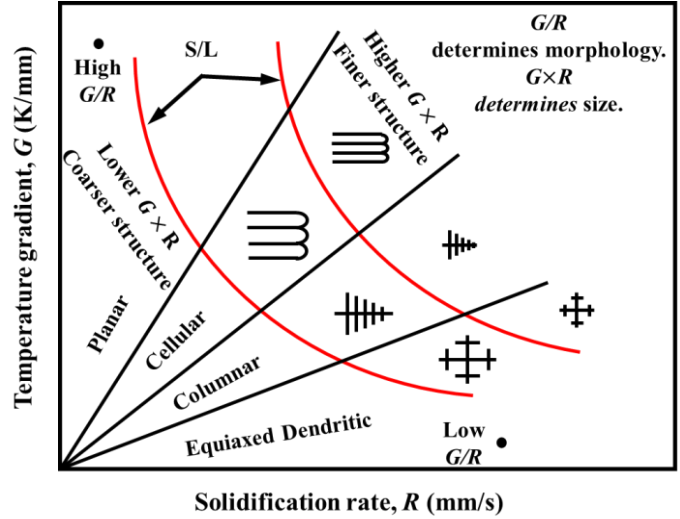

Fig. 14. Effects of $\boldsymbol{G}$ and $\boldsymbol{R}$ on the morphology and size of the solidified microstructure in laser cladding process [12].

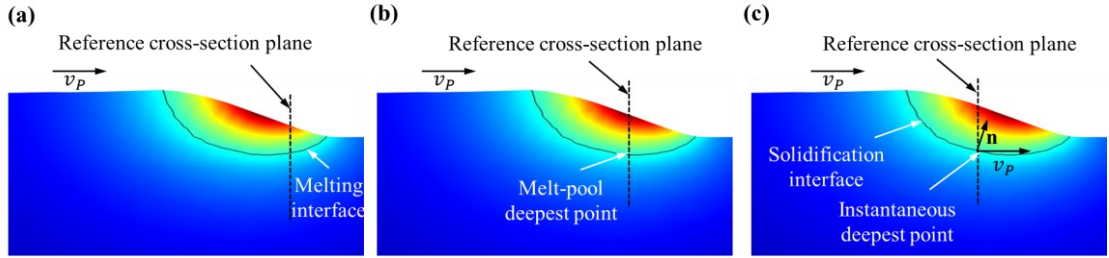

Fig. 15. Molten pool passes through the fixed reference cross-section: (a) before passing, (b) on passing, (c) after passing.

To predict the effect of temperature gradient $\boldsymbol{G}$ and solidification rate $\boldsymbol{R}$ on the microstructure evolution, the values of $\boldsymbol{G}, \boldsymbol{R}$ at the instantaneous deepest point of solidification interface in the reference cross-section plane (as shown in Fig. 15) for three time instances, $t$=0.38 s, 0.40 s and 0.42 s was calculated. The corresponding points locate on the top region, interior region, and bottom region of the sample bead cross-section as marked in Fig. 7. The cladding process parameters are as follows: laser power 56 W, powder feed rate 0.83 g/min, and scan speed 2 mm/s. The simulated temperature gradient $\boldsymbol{G}$, solidification rate $\boldsymbol{R}$ in the x, y, z directions as well as the shape factor of $\boldsymbol{G}/\boldsymbol{R}$ are presented in Table 4. As shown in Table 4, the value of $\boldsymbol{G}$ on the three points has no significant difference, the maximum value of temperature gradient magnitude is 4192 K/mm at the bottom point 'C' while the minimum value is 2871 K/mm at the top point 'A'. However, the difference in $\boldsymbol{R}$ is significant as the normal vector of solidification front $\boldsymbol{n}$ varies greatly. At the bottom point 'C', due to the smallest of melt pool retreat force, the normal vector is almost in z direction, yielding a small solidification rate $R_z = 0.73$ mm/s in the z-direction. At the top point 'A', the normal vector is approximately in the scan direction, resulting in a large solidification rate $R_z = 6.15$ mm/s in the z-direction. The solidification rate $R_y$ in the y-direction is kept constant (the velocity is equal to the scan speed) for the all three points, due to the molten pool quasi-steady state.

**Table 4**
Temperature gradient, solidification rate, and the shape factor at points A, B, C in Fig. 7.

| Point | Location | Temperature gradient $G$ (K/mm) | Solidification rate $R$ (mm/s) | Shape factor $G/R$ (s*K/mm²) |
|---|---|---|---|---|
| A | Refer to Fig. 7 | $\begin{bmatrix} G_x \\ G_y \\ G_z \end{bmatrix} = - \begin{bmatrix} 0 \\ 2675 \\ 1043 \end{bmatrix}$ | $\begin{bmatrix} R_x \\ R_y \\ R_z \end{bmatrix} = \begin{bmatrix} 0 \\ 2 \\ 6.15 \end{bmatrix}$ | 443 |
| B | Refer to Fig. 7 | $\begin{bmatrix} G_x \\ G_y \\ G_z \end{bmatrix} = - \begin{bmatrix} 0 \\ 3062 \\ 2251 \end{bmatrix}$ | $\begin{bmatrix} R_x \\ R_y \\ R_z \end{bmatrix} = \begin{bmatrix} 0 \\ 2 \\ 3.20 \end{bmatrix}$ | 1007 |
| C | Refer to Fig. 7 | $\begin{bmatrix} G_x \\ G_y \\ G_z \end{bmatrix} = - \begin{bmatrix} 0 \\ 524 \\ 4160 \end{bmatrix}$ | $\begin{bmatrix} R_x \\ R_y \\ R_z \end{bmatrix} = \begin{bmatrix} 0 \\ 2 \\ 0.73 \end{bmatrix}$ | 1970 |

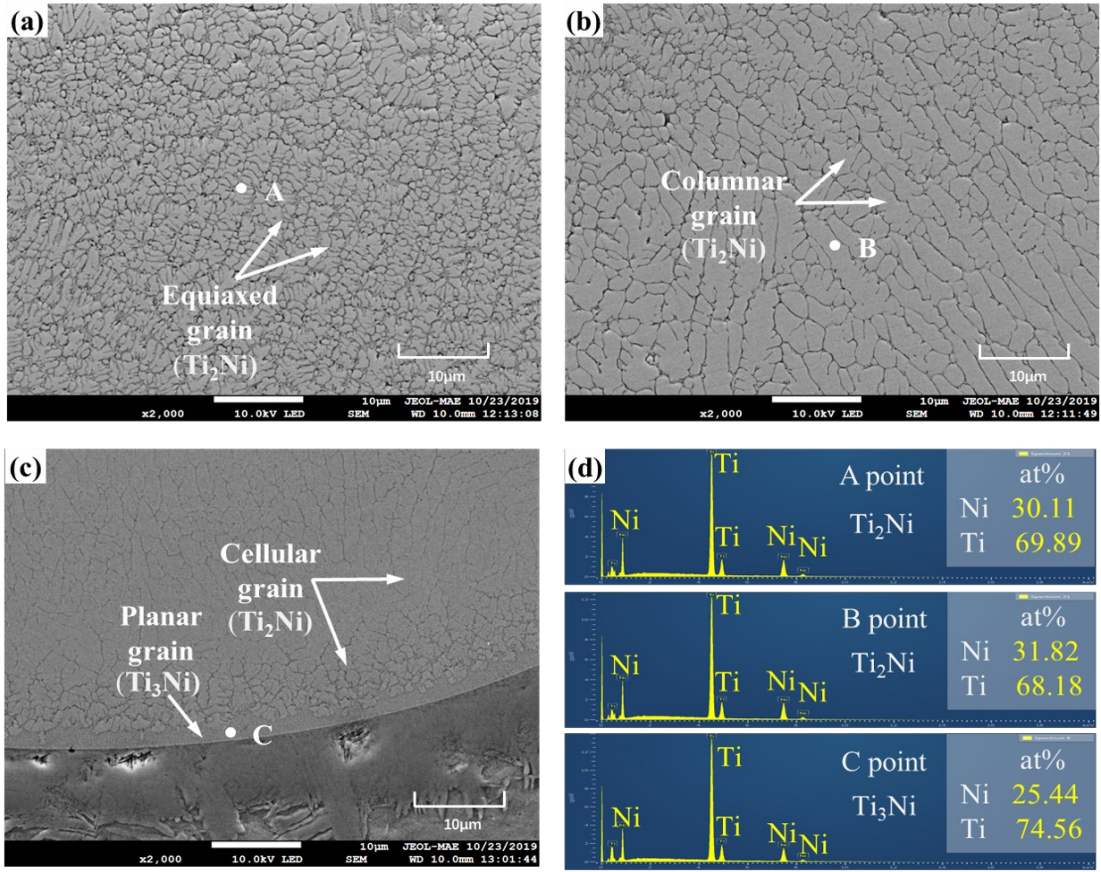

Fig. 16. The microstructures at the points of (a) A-top, (b) B-middle, (c) C-bottom; and (d) their corresponding element distribution.

For well-mixed samples, the typical microstructure characteristics of a single track at these three different locations in a cross-section are shown in Fig. 16. The grains of the track from the bottom to the top experience planar, cellular, columnar, and equiaxed grains are shown. In the initial stage of the molten pool solidification process, the temperature gradient $G$ at the bottom is very high but the solidification rate $R$ at the same point is relatively low, which results in the shape factor $G/R$ at the bounding interface extremely high. Moreover, large amounts of Ti elements exist in the liquid-solid boundary. The combined effect of both induces the $Ti_3Ni$ planar grain growth on the solidification interface as shown in Fig. 16(c) and Fig. 16(d). Along with the solidification progresses, the newly formed planar grains will impede the heat exchange between the substrate and the molten pool, decreasing the temperature gradient *G* on the area near the solidification interface. Meanwhile, the direction of heat flow between the molten pool and the substrate is perpendicular to the solidification interface and the solidification rate *R* slowly increases. With these conditions, the cellular grains perpendicular to the previously formed planar grain zone are gradually formed. When the movement of the solidification front towards the surface, the temperature gradient *G* is further decreased, leading to the increment of undercooling. This undercooling will also be further aggravated by the formation of new cellular grains, accelerating the secondary nucleation in the liquid metal. Therefore, the $Ti_2Ni$ columnar dendrites are formed as shown in Fig. 16(b). At the top area, the temperature gradient *G* is lowest due to the poor heat dissipation condition. This lower *G* is beneficial to the nucleation rate. The solidification rate *R* at this area is extremely high, which further reduces the time of grain growth. The increased nucleation rate, as well as decreased growth time, induces the formation of $Ti_2Ni$ equiaxed grains on the top surface in Fig. 16(a).

## 5. Conclusion

To investigate the directed energy deposition of NiTi shape memory alloy, a 3D transient multiphase model considering heat transfer and fluid flow was proposed. The developed model can be used to predict the width, height, depth, deposition rate of the single-track and to investigate the effect of fluid velocity, temperature gradient *G*, solidification rate *R*, shape factor *G/R* on the microstructure morphology. A series of single-track experiments were then conducted under the following process conditions: laser power changes from 45 W to 78.7 W, laser scan speed varies from 2 mm/s to 4 mm/s, and powder feed rate increases from 0.69 g/min to 0.96 g/min to validate the model. The following conclusions can be obtained from the simulated and experimental results:

(1) The clad width is primarily determined by the laser power while the clad height is mainly influenced by the powder density per unit area. The clad depth is jointly determined by laser power energy, powder feeding rate, and scan speed.
(2) The deposition rate is significantly affected by the powder feed rate and laser power energy, the scan speed has little effect on it.
(3) The distribution of elements in the molten pool is affected by fluid velocity, the process parameters have a huge influence on it.
(4) The grain morphology is mainly affected by the temperature gradient *G* and the solidification rate *R*. The shape factor *G /R* makes the molten pool present diverse grain types along with the solidification interface.

In our future work, element evaporation and transportation, microstructure evolution, and secondary phase precipitation will be simulated to fully realize the phase transformation temperature prediction for the NiTi shape memory alloy.

## Acknowledgments

This research was supported by the start-up fund from McCormick School of Engineering, Northwestern University, Evanston, IL, USA; the Research Grants Council of the Hong Kong Special Administrative Region, China (Project No. CUHK 14202219) and The Chinese University of Hong Kong